\documentclass[a4paper,UKenglish]{lipics-v2016}
\newcommand{\ceil}[1]{\left \lceil #1 \right \rceil}

\usepackage{microtype}%if unwanted, comment out or use option "draft"

\title{Succinct Representation for (Non)Deterministic Finite Automata}
\titlerunning{Succinct Representation for (Non)Deterministic Finite Automata} %optional, in case that the title is too long; the running title should fit into the top page column

\author[1]{Sankardeep Chakraborty}
\author[2]{Roberto Grossi}
\author[3]{Kunihiko Sadakane}
\author[4]{Srinivasa Rao Satti}
\affil[1]{RIKEN Center for Advanced Intelligence Project, Japan\\
  \texttt{sankar.chakraborty@riken.jp}}
\affil[2]{Dipartimento di Informatica, Università di Pisa, Italy\\
  \texttt{grossi@di.unipi.it}}
  \affil[3]{The University of Tokyo, Japan\\
  \texttt{sada@mist.i.u-tokyo.ac.jp}}
  \affil[4]{Seoul National University, South Korea\\
  \texttt{ssrao@cse.snu.ac.kr}}
\authorrunning{Chakraborty, Grossi, Sadakane and Satti} %mandatory. First: Use abbreviated first/middle names. Second (only in severe cases): Use first author plus 'et. al.'

\Copyright{Chakraborty, Grossi, Sadakane and Satti}%mandatory, please use full first names. LIPIcs license is "CC-BY";  http://creativecommons.org/licenses/by/3.0/

\subjclass{Dummy classification -- please refer to \url{http://www.acm.org/about/class/ccs98-html}}% mandatory: Please choose ACM 1998 classifications from http://www.acm.org/about/class/ccs98-html . E.g., cite as "F.1.1 Models of Computation". 
% mandatory: Please provide 1-5 keywords
\EventEditors{John Q. Open and Joan R. Acces}
\EventNoEds{2}
\EventLongTitle{42nd Conference on Very Important Topics (CVIT 2016)}
\EventShortTitle{CVIT 2016}
\EventAcronym{CVIT}
\EventYear{2016}
\EventDate{December 24--27, 2016}
\EventLocation{Little Whinging, United Kingdom}
\EventLogo{}
\SeriesVolume{42}
\ArticleNo{23}
\begin{document}

\maketitle

\begin{abstract}
Deterministic finite automata are one of the simplest and most practical models of computation studied in automata theory. Their conceptual extension is the non-deterministic finite automata which also have plenty of applications. In this article, we study these models through the lens of succinct data structures where our ultimate goal is to encode these mathematical objects using information theoretically optimal number of bits along with supporting queries on them efficiently. Towards this goal, we first design a succinct data structure for representing any deterministic finite automaton $\mathcal{D}$ having $n$ states over a $\sigma$-letter alphabet $\Sigma$ using $(\sigma-1) n\log n + O(n \log \sigma)$
%$(\sigma-1) \lceil n\log n\rceil + O(n \log \sigma)$
%$(\sigma-1)n (\log n + O(1))+O(n \log\sigma)+3n+o(n)$ 
bits of space, which can determine, given an input string $x$ over $\Sigma$, whether $\mathcal{D}$ accepts 
$x$ in $O(|x| \log \sigma)$ time, using constant words of working space. 
When the input deterministic finite automaton is acyclic, not only we can improve the above space bound significantly to $(\sigma -1) (n-1)\log n+ 3n + O(\log^2 \sigma) + o(n)$ bits, we also obtain optimal query time for string acceptance checking. More specifically, using our succinct representation, we can check if a given input string $x$ can be accepted by the acyclic deterministic finite automaton using time proportional to the length of $x$, hence, the optimal query time.
%$(\sigma -1) \lceil n\log n\rceil + 3n + o(n)$ 
We also exhibit a succinct data structure for representing a non-deterministic finite automaton $\mathcal{N}$ having $n$ states over a $\sigma$-letter alphabet $\Sigma$ using $\sigma n^2+n$ bits of space, such that given an input string $x$, we can decide whether $\mathcal{N}$ accepts $x$ efficiently in $O(n^2|x|)$ time. Finally, we also provide time and space efficient algorithms for performing several standard operations such as union, intersection and complement on the languages accepted by deterministic finite automata. 
\end{abstract}

\section{Introduction}
\begin{figure}[htp]
	\begin{center}
		\includegraphics[scale=0.4]{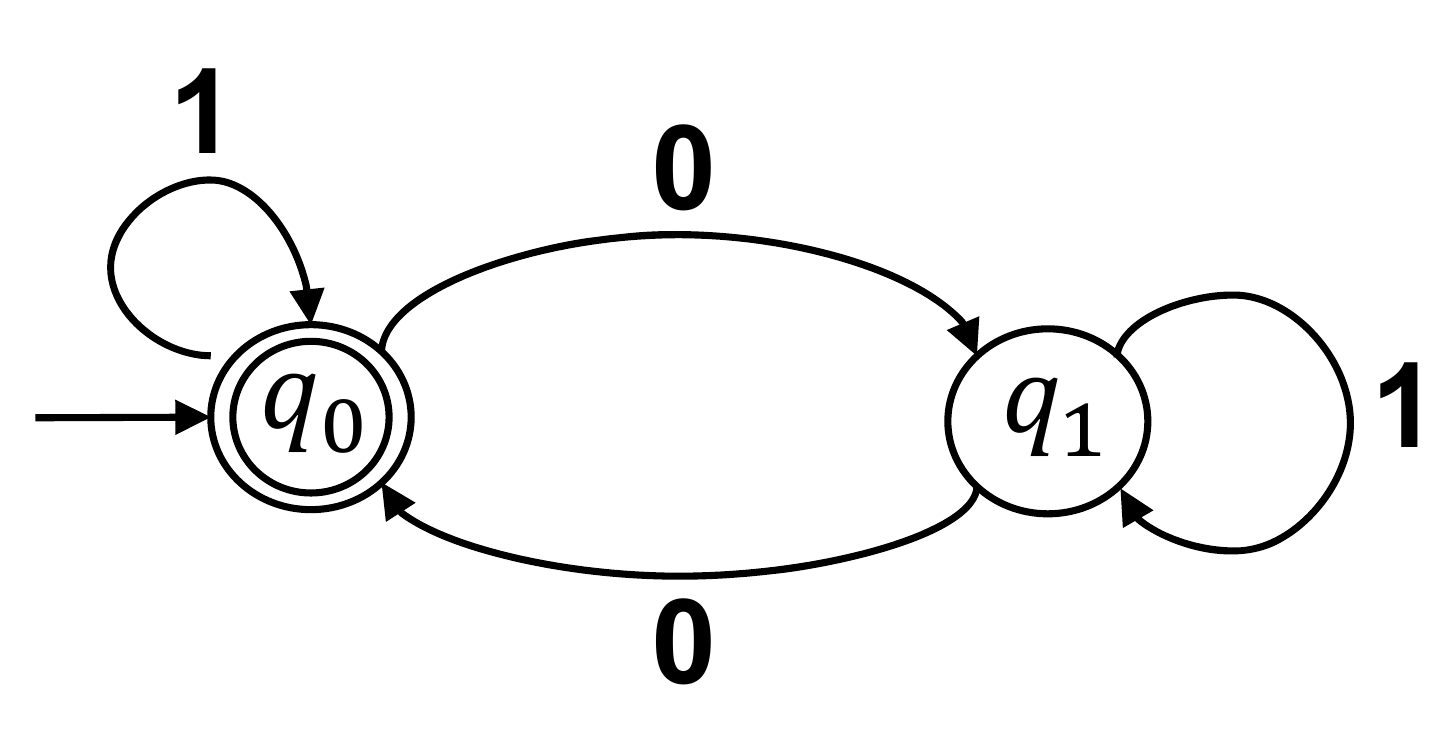}
	\end{center}
	\caption{The {\it state transition diagram} for a DFA $\mathcal{D}$ where $\mathcal{D} =(\Sigma, Q, q_0, \delta, F)$ such that (i) $\Sigma=\{0,1\}$, (ii) $Q=\{q_0,q_1\}$, (iii) $q_0=q_0$ (marked with an incoming arrow coming from nowhere), (iv) $F=\{q_0\}$, and (v) the transition function is defined as the following set,
$\{\delta(q_0,1)=q_0,\delta(q_0,0)=q_1, \delta(q_1,1)=q_1,\delta(q_1,0)=q_0\}$. Precisely the DFA $\mathcal{D}$ accepts all the strings containing an even number of zeros over the binary alphabet.}
	\label{figure:DFA}
\end{figure}

Automata theory is a branch of theoretical computer science that deals exclusively with the definitions, properties and applications of different mathematical models of computation. These models play a major role in multiple applied areas of computer science. One of the most basic and fundamental models that is studied in automata theory since long time back is called the {\it finite automata}. It primarily comes in two different types, {\it deterministic finite automata} (henceforth DFA) and {\it non-deterministic finite automata} (henceforth NFA) among others. There exists more complex and sophisticated  models as well, for example, {\it Context-free grammar, Turing machines} etc. In what follows, let us formally define DFA and NFA in a nutshell as these are our primary subjects of study in this article. A DFA $\mathcal{D}$ is a quintuple $\mathcal{D}=(\Sigma, Q, q_0, \delta, F)$  where:  
\begin{itemize}
\item $\Sigma$ is an {\it alphabet}; a finite set of letters,
\item $Q$ is the finite set of {\it states},
\item $q_0 \in Q$ is the {\it initial state},
\item $\delta: Q \times \Sigma \rightarrow Q$ is the {\it transition function} and
\item $F \subseteq Q$ is the {\it set of final states}. 
\end{itemize}
We often extend the transition function to $\delta: Q \times \Sigma^* \rightarrow Q$ which is defined recursively as follows: $\delta(q, \epsilon) = q$ for all $q \in Q$, where $\epsilon$ is the empty string; and $\delta(q, aw) = \delta(\delta(q,a),w)$ for all $q \in Q$, $a \in \Sigma$, and $w \in \Sigma^*$.
Given the above definition, we say that the DFA accepts a string $x$ over the alphabet $\Sigma$ if and only if $\delta(q,x) \in F$.
%Then we say that the DFA $\mathcal{D}$ {\it accepts} the string $x$ if and only if there exists a sequence of states $y_0,y_1, y_2, \ldots, y_n$ in $Q$ such that (i) $y_0=q_0$, (ii) $y_{i+1}=\delta(y_i,x_{i+1})$, for $i=0,1, \ldots,n-1$, and (iii) $y_n \in F$. 
The {\it language} $\mathcal{L}$ accepted by a DFA $\mathcal{D}$ is defined as the set of all strings accepted by the DFA $\mathcal{D}$, and is denoted by $\mathcal{L}(\mathcal{D})$. See Figure~\ref{figure:DFA} for a simple example.
In the rest of this paper, we assume that the alphabet $\Sigma$ is $\{1, 2, \dots, \sigma \}$, and the state set $Q$
is $\{ q_0, q_1, \dots, q_{n-1} \}$.

A deterministic automaton $\mathcal{A}$ is called {\it acyclic}~\cite{Liskovets06} if it has a unique recurrent state where a state $q$ is defined as {\it recurrent} if there exists a non-empty string $x$ over $\Sigma$ such that $\delta(q,x)=q$. Non-recurrent states are typically called {\it transient}, and the unique recurrent state (denoted by $q'' \in Q$) is classically called the {\it dead state} as $\delta(q'',\sigma)=q''$ for all $\sigma \in \Sigma$. 

An NFA is a conceptual extension of DFAs where the definition of the transition function is mainly extended. More specifically, for DFA, the transition function is defined as $\delta$: $Q \times \Sigma \rightarrow Q$ whereas for NFA, the same is defined as $\delta$: $Q \times \Sigma \rightarrow \mathcal{P}(Q)$ where $\mathcal{P}(Q)$ denotes the power set of $Q$. Another extension, which is sometimes used in the literature, is to simply allow more than one initial state in an NFA, and in this case, the third item in the tuple becomes $I$ denoting the set of initial states, instead of singleton $\{q_0\}$. The rest of above quintuple definition remains as it is for NFA. Thus, in the case of NFA $\mathcal{N}$, the language $\mathcal{L(N)}$ is defined as $\{x \mid \exists_{q\in I} \exists_{q' \in F} [q' \in \delta(q,x)]\}$. We refer the readers to the classic texts of~\cite{Hopcroft,Sipser} for a thorough discussions on these mathematical models and automata theory in general. 

Even if a DFA is defined as an abstract mathematical concept, still it has got myriad of practical applications. More specifically, it is used in text processing, compilers, and hardware design~\cite{Sipser}. Quite often it is implemented in small hardware and software tools for solving various specific tasks. For example, a DFA can model a software that can figure out whether or not online user input such as email addresses are valid.
DFAs/NFAs are also used for network packet filtering.
In some of these applications, the alphabet is large and there is a failure/exit state so that only a subset of transitions go to non-failure states; so we call the latter ones \emph{non-failure} transitions. 

Despite having so many applications in practically motivated problems, we are not aware of, to the best of our knowledge, any study of DFAs and NFAs from the point of view of \textit{succinct data structures} where the goal is to store an arbitrary element from a set $Z$ of objects using the information theoretic minimum $\log(|Z|)+o(\log(|Z|))$ bits of space while still being able to support the relevant set of queries efficiently, which is what we focus on in this paper. We also assume the usual model of computation, namely a $\Theta (\log n)$-bit word RAM model where $n$ is the size of the input. 

\subsection{Related Work}
The field of {\it succinct data structures} originally started with the work of Jacobson~\cite{Jacobson}, and by now it is a relatively mature field in terms of breadth of problems considered. To illustrate this further, there already exists a large body of work on representing various combinatorial objects succinctly. A partial list of such combinatorial objects would be trees~\cite{MunroR01,NavarroS14}, various special graph classes like planar graphs~\cite{AleardiDS08}, chordal graphs~\cite{MunroW18}, partial $k$-trees~\cite{FarzanK14}, interval graphs~\cite{Acan} along with arbitrary general graphs~\cite{FarzanM13}, permutations~\cite{MunroRRR12}, functions~\cite{MunroRRR12}, bitvectors~\cite{RamanRS07} among many others. We refer the reader to the recent book by Navarro~\cite{Navarro} for a comprehensive treatment of this field. The study of succinct data structures is motivated by both theoretical curiosity and also by the practical needs as these combinatorial structures do arise quite often in various applications.  

For DFA and NFA, other than the basic structure that is mentioned in the introduction, there exists many extensions/variations in the literature, for example, two-way finite automata, B\"{u}chi automata and many more. Researchers generally study the properties, limitations and applications of these mathematical structures. One such line of study that is particularly relevant to us for this paper is the research on counting DFAs and NFAs. Since the fifties there are plenty of attempts in exactly counting the number of DFAs and NFAs with $n$ states over the alphabet $\Sigma$, and the state-of-the-art result is due to~\cite{BassinoN07} for DFAs and~\cite{DomaratzkiKS02} for NFAs respectively. We refer the readers to the survery (and the references therein) of Domaratzki~\cite{Domaratzki06} for more details. Basically, from these results, we can deduce the information theoretic lower bounds on the number of bits required to represent any DFA or NFA. Then we augment these lower bounds by designing data structures whose size matches the lower bounds, hence consuming optimal space, along with capable of executing algorithms efficiently using this succinct representation, and this is the main contribution of this paper.

\subsection{DFA and NFA Enumeration}\label{enumeration}
After a number of efforts by several authors, finally Bassino and Nicaud~\cite{BassinoN07} found a matching upper and lower bound on the number of non-isomorphic initially-connected\footnote{Note that this assumption always implies that the language accepted by the DFA is non-empty.} (i.e., all the states are reachable from the initial state) DFA's with $n$ (including a fixed initial and one or possibly more final) states over an alphabet $\Sigma$ (where $|\Sigma|=\sigma$) is $\Theta(n2^{2n}S_2(\sigma n,n))$ where $S_2(n,m)$ denotes the Stirling numbers of the second kind\footnote{It is defined recursively as $S_2(0,0)=1$, $S_2(n,0)=0$ for all $n \geq 1$ and for all $n,m \geq 1$, $S_2(n,m)=m S_2(n-1,m)+S_2(n-1,m-1)$.}.
Using the approximation of the Stirling numbers of the second kind~\cite{Flajolet}, which states that $S_2(n,m) \thickapprox \frac{m^n}{m!}$, we can obtain the information theoretic lower bound for representing any DFA having $n$ states and $\sigma$-sized alphabet is given by $\lg (n2^{2n}S_2(\sigma n,n))=(\sigma -1) n \lg n+ O(n)$ bits. On the other hand, Domaratzki et al.~\cite{DomaratzkiKS02} showed that there are asymptotically $2^{\sigma n^2+n}$ initially connected NFAs on $n$ states over a $\sigma$-letter alphabet with a fixed initial state and one or more final states. Thus, information theoretically, we need at least $\sigma n^2+n$ bits to represent any NFA. In what follows later, we show that we can represent any given DFA/NFA using asymptotically optimal number of bits as mentioned here. Throughout this paper, we assume that the input DFAs/NFAs that we want to encode
succinctly are initially connected.

\subsection{Our Main Results and Paper Organization}
The classical representation of DFAs/NFAs consists of explicitly writing the transition function $\delta$ in a two dimensional array $J[0..n-1][1..\sigma]$ having $n$ rows corresponding to the $n$ states of the DFA/NFA and $\sigma$ (where $|\Sigma|=\sigma$) columns corresponding to the alphabet $\Sigma$ such that $J[i][j]=\delta(q_i,j)$ where $q_i \in Q, j \in \Sigma$. For DFA, the entry in $J[i][j]$ is a singleton set whereas for NFA it could possibly contain a set having more than one state. Thus, the space requirement for representing any given DFA (NFA respectively) is given by $O(n \sigma \log n)$ ($O(n^2 \sigma \log n)$ respectively) bits. These space bounds are clearly not optimal -- for the DFAs, it is off by an additive $n \log n$ term from the information theoretic minimum, while for the NFAs, it is off by a multiplicative factor of $ \log n$ from the optimal bound. We alleviate this discrepancy in the space bounds by designing optimal succinct data structures for these objects. 

Towards this goal, we start by listing all the preliminary data structures and graph theoretic terminologies that will be required in our paper in Section~\ref{prelims}. Then, in Section~\ref{dfa_encoding} we first discuss the relevant prior work from~\cite{BassinoN07}, and show that, by using suitable data structures, their work already gives a succinct encoding of DFA. But the major drawback of this encoding is that it is not capable of handling the problem of checking whether a string is accepted by the DFA extremely efficiently. In Section~\ref{dfa_succinct}, we overcome this problem by designing a succinct data structure for DFA, which can also check the string acceptance almost optimally. We summarize our main result in the following theorem.

\begin{theorem}\label{dfa_main_result}
Given an initially-connected deterministic finite automata $\mathcal{D}$ having $n$ states and working over an alphabet $\Sigma$ of size $\sigma$, there exists a succinct encoding for $\mathcal{D}$ taking $(\sigma-1) n\log n + O(n \log \sigma)$
%$(\sigma-1) \lceil n\log n\rceil + O(n \log \sigma)$
%$(\sigma-1)n (\log n + O(1))+O(n \log\sigma)+3n+o(n)$ 
bits of space, which can determine, given an input string $x$ over $\Sigma$, whether $\mathcal{D}$ accepts $x$ in 
$O(|x| \log \sigma)$ time, using constant words of working space.
If the DFA has only $N < \sigma n$ non-failure transitions, then the space can be further reduced to %sada $N \log(\sigma n^2/ N) + o(N)$ bits.
$(N-n) \log n + O(N \log\sigma)$ bits.
\end{theorem}

The upper bounds in Theorem~\ref{dfa_main_result} save roughly $n \log n$ bits with respect to the immediate representation of the DFA. The former upper bound is optimal as it matches the information-theoretical lower bound in Section~\ref{enumeration}, up to lower order terms. As for the latter upper bound, we do not know its optimality but it is smaller than the information-theoretical lower bound of $\lceil \log {n^2 \choose N} \rceil + \Theta(N \log \sigma)$ bits derived for edge-labeled deterministic directed graphs~\cite{FarzanM13}. Indeed, DFAs can be seen as a special case of these graphs where $n$ is the number of nodes, $N \geq n-1$ is the number of arcs, and $\sigma$ is the maximum node degree.\footnote{A directed graph with labels on its arcs is deterministic if no two out-neighbor arcs have the same label. Since there are $\lceil \log {n^2 \choose N} \rceil$ directed graphs~\cite{FarzanM13} with $n$ nodes and $N$ arcs, each deterministic graph $G=(V,E)$ can have $L = \prod_{u \in V} d_u!$ label assignments for its arcs, where $d_u$ s the out-degree of node $u$ and $N = \sum_{u \in V} d_u$. Note that $\log L = \Theta(N \log \sigma)$ when labels are from $\Sigma$ and thus $d_u \leq \sigma$.}

We can improve the above space bound significantly if the given DFA is acyclic along with obtaining optimal query time for string acceptance checking. More specifically, in Section~\ref{acyclic_dfa_succinct}, we obtain the following result in this case.

\begin{theorem}\label{acyclic_dfa_main_result}
Given an initially-connected acyclic deterministic finite automata $\mathcal{A}$ having $n-1$ transient states, a unique dead state and working over an alphabet $\Sigma$ of size $\sigma$, there exists a succinct encoding for $\mathcal{A}$ taking
$(\sigma -1) (n-1)\log n+ 3n + O(\log^2 \sigma) + o(n)$  
%$(\sigma-1) \lceil n\log n\rceil + 2n+o(n)$
%$(\sigma-1)n (\log n + O(1))+O(n \log\sigma)+3n+o(n)$ 
bits of space, which can optimally determine, given an input string $x$ over $\Sigma$, whether $\mathcal{A}$ accepts $x$ in time proportional to the length of $x$, using constant words of working space.
\end{theorem}

This is followed by the succinct data structure for NFA in Section~\ref{nfa_succinct} where we prove the following result.

\begin{theorem}\label{nfa_main_result}
Given an initially-connected non-deterministic finite automata $\mathcal{N}$ having $n$ states and working over an alphabet $\Sigma$ of size $\sigma$, there exists a succinct encoding for $\mathcal{N}$ taking $\sigma n^2+n$ bits of space, which can determine, given an input string $x$ over $\Sigma$, whether $\mathcal{N}$ accepts $x$ in $O(n^2|x|)$ time, using $2n$ bits of working space.
\end{theorem}

Next we move on to discuss how one can support several standard operations such as union and intersection of two languages accepted by the deterministic finite automata. Classically it is done via the product automaton construction~\cite{Hopcroft,Sipser}, and here we provide a time and space efficient algorithm for performing this construction. More specifically, we show the following theorem (proof and other details are provided in Appendix~\ref{product_construction}),

\begin{theorem}\label{thm:product}
Suppose we are given the succinct representations for two DFAs $\mathcal{D}_1$ (having $n$ states) and $\mathcal{D}_2$ (having $n'$ states) respectively such that both are working over the same alphabet $\Sigma$. Also suppose that the product automata (denoted by $\mathcal{P}$) has $n''$ states where $n'' \leq nn'$. Then, using $O(n'')$ expected time and $O(n'' \log n'')$ bits of working space, we can directly construct a succinct representation for $\mathcal{P}$. Moreover, $\mathcal{P}$ can be represented optimally using $(\sigma-1) n'' \log n''+O(n'' \log\sigma)$ bits overall, and by suitably defining the final states of $\mathcal{P}$, we can make $\mathcal{P}$ accept either $\mathcal{L}(\mathcal{D}_1) \cup \mathcal{L}(\mathcal{D}_2)$ or $\mathcal{L}(\mathcal{D}_1) \cap \mathcal{L}(\mathcal{D}_2)$. Finally, given an input string $x$ over $\Sigma$, we can decide whether $x \in \mathcal{L}(\mathcal{P})$ in $O(|x| \log \sigma)$ time using constant words of working space. 
\end{theorem}

Finally, we conclude in Section~\ref{conclusion} with some concluding remarks.

\section{Preliminaries}\label{prelims}
In this section we collect all the previous theorems and definitions that will be used throughout this paper. 
\subsection{Graph Terminology and Graph Algorithms}
We will assume the knowledge of basic graph theoretic terminology (like trees, paths etc) as given in~\cite{Diestel} and basic graph algorithms (mostly the depth first search (henceforth DFS) traversal of a graph and its related concepts) as given in~\cite{CLRS}. Perhaps at this point it may seem slightly unusual that we are talking about graphs here when the focus of this paper is DFA/NFA and their succinct representations. Essentially in this paper we view DFA/NFA, more specifically their graphical representation i.e., {\it state transition diagram}, as a special case of an edge labeled directed graph $G$ having $n$ nodes corresponding to the $n=|Q|$ states of DFA/NFA, $m=\sigma n$ edges where $|\Sigma|=\sigma$ as each node has exactly $\sigma$ outgoing edges, and each edge is labeled with some elements from $\Sigma$. It is with this point of view, we will design our succinct data structures for DFA/NFA in this paper.

\subsection{Succinct Data Structures}\label{prelims_stricture}
{\bf Rank-Select.} 
For a bit vector $B$ and any $ a\in \{0,1\}$, the rank and select operations are defined as follows :
\begin{itemize}
 \item $rank_a(B,i)$ = the number of occurrences of $a$
 %$a\in \{0,1\}$ 
 in $B[1,i]$, for $1\leq i\leq n$;
 \item $partial\_rank_1(B,i)$ = $rank_1(B,i)$ if $B[i] = 1$, and $-1$ otherwise; and
 \item $select_a(B,i)$ = the position in $B$ of the $i$-th occurrence of $a$, for $1\leq i\leq n$.
% $a\in \{0,1\}$.
\end{itemize}

We make use of the following theorems:
\begin{theorem}~\cite{Clark96}
 \label{staticrs}
We can store a bitstring $B$ of length $n$ with additional $o(n)$ bits such that rank and select operations can be supported in $O(1)$ time. Such a structure can also be constructed from the given bitstring in $O(n)$ time and space.
\end{theorem}

\begin{theorem}~\cite{RamanRS07}
 \label{thm:id}
We can store a bitstring $B$ of length $n$ with $m$ ones using $\log{n \choose m} + o(m) +O(\log\log n)$ bits such that $partial\_rank_1$ operations can be supported in $O(1)$ time. Such a structure can also be constructed from the given bitstring in $O(n)$ time and space.
\end{theorem}

\noindent
{\bf Succinct tree representation.} We use following result from~\cite{MunroR01}.
\begin{theorem}~\cite{MunroR01}
\label{succ_tree}
Given a rooted ordered tree $\tau$ on $n$ nodes, it can be succinctly  represented as a sequence of balanced parenthesis of length $2n$ bits, such that given a node $v$, we can support subtree size and various navigational queries (such as parent and $i$-th child) on $v$ in $O(1)$ time using an additional $o(n)$ bits. Such a structure can also be constructed in $O(n)$ time and space.
\end{theorem}

\noindent
{\bf Compact representation of increasing sequence.} We use the following theorem from~\cite{Sumigawa18}.

\begin{theorem}~\cite{Sumigawa18}
\label{increasing}
Given an increasing integer sequence $a[\cdot]$ of length $n$ such that $0 \le a[1] \le a[2] \le \cdots \le a[n] < u$, there exists a data structure to represent $a[\cdot]$ in compressed form using $O(\min\{\frac{1}{\epsilon}n^\epsilon u^{1-\epsilon}, \frac{1}{\epsilon}u^\epsilon n^{1-\epsilon}\})$ bits of space, where $\epsilon>0$ is any parameter, such that any entry $a[i]$ and the value 
%$\overline{a}[i] = |\{j \mid a[j] > i, 1 \le j \le n\}|$
$\overline{a}[i] = |\{j \mid a[j] < i, 1 \le j \le n\}|$
can be retrieved in $O(1/\epsilon)$ time. 
\end{theorem}
We denote the above data structure by $D(n, u, \epsilon)$.
If $B$ denotes the characteristic vector for the sequence $a$, then
computing $a[i]$ and $\overline{a}[i]$ correspond to computing 
select and rank on $B$.

%Given an increasing integer sequence $a[\cdot]$ of length $n$ such that $0 \le a[1] \le a[2] \le \cdots \le a[n] < u$, there exists a data structure to represent $a[\cdot]$ in compressed form using $O(\min\{\frac{1}{\epsilon}n^\epsilon u^{1-\epsilon}, \frac{1}{\epsilon}u^\epsilon n^{1-\epsilon}\})$ bits of space, where $\epsilon>0$ is any fixed constant, such that any entry $a[i]$ can be still retrieved in constant time. We denote it by $S(n, u)$.
%\end{theorem}

\medskip
\noindent
{\bf Representation of a vector.} We also make use of the following theorem from~\cite{DodisPT10}.

\begin{theorem}~\cite{DodisPT10}
\label{vector_representation}
There exists a data structure that can represent a vector $A[1..n]$ of elements from a finite alphabet $\Sigma$ using 
$n \log |\Sigma|  + O(\log^2 n)$ 
%$\lceil n \log \Sigma \rceil + O(\log^2 n)$ 
bits, such that any element of the vector can be read or written in constant time. 
\end{theorem}

\section{Succinct Representations for DFA and NFA}
In this section, we provide all the upper bound results of our paper dealing with DFA/NFA. Throughout this section, whenever we mention DFA (NFA resp.), it should refer to an initially-connected deterministic (non-deterministic resp.) finite automata having $n$ states and working over an alphabet $\Sigma$ of size $\sigma$. With this notation in mind, we start with the succinct encoding of DFA first.

\subsection{Succinct Encoding of DFA}\label{dfa_encoding}
\newcommand{\MAX}{{\it Max}}
\newcommand{\BOXED}{{\it Boxed}}
\newcommand{\NBOXED}{{\it NewBoxed}}

\begin{figure}[htp]
	\begin{center}
		\includegraphics[scale=0.4]{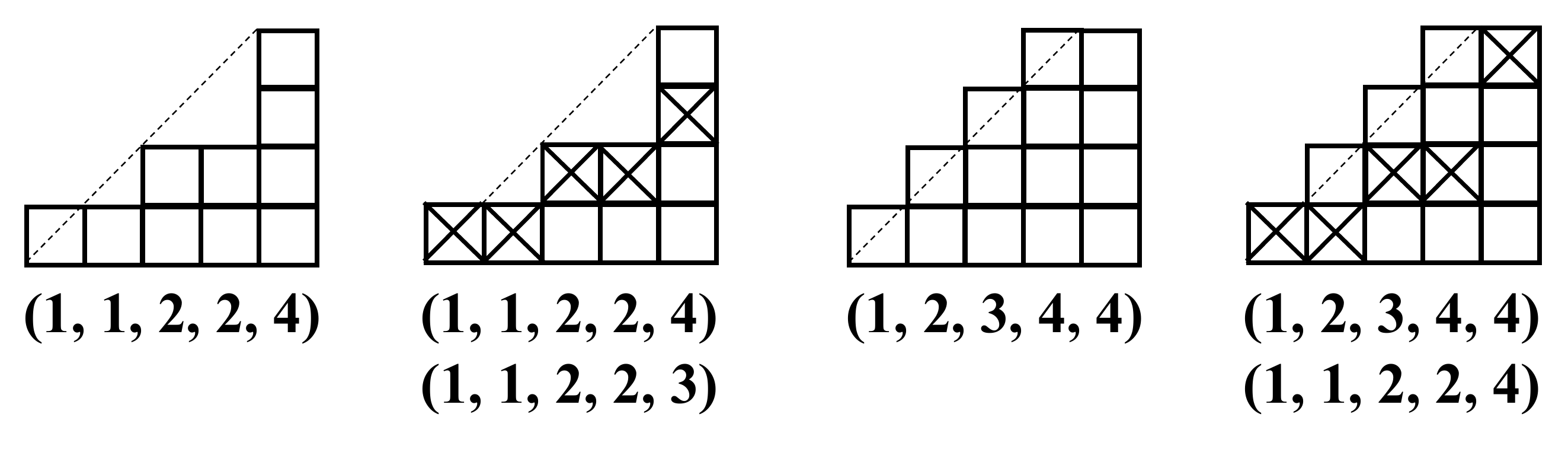}
	\end{center}
	\caption{A diagram of width $m=5$ and height $n=4$, a boxed diagram, a $k$-Dyck diagram and a $k$-Dyck boxed diagram with $k=2$.}
	\label{boxandmatch}
\end{figure}

Bassino and Nicaud~\cite{BassinoN07} proved a beautiful bijection between the state transition diagram of any DFA and pairs of integer sequences which can be represented by boxed diagrams (will be defined shortly) along with providing an efficient algorithm to perform this construction. We will refer the readers to~\cite{BassinoN07}
%their beautifully written paper 
for complete details regarding the bijection, counting and many other details that we choose to not repeat here. However, we still need to provide some details/definitions (which basically follow their exposition) that are relevant to our own work and will also help to understand the results from their paper smoothly. Following~\cite{BassinoN07}, a {\it diagram} of width $m$ and height $n$ is defined as a sequence $(x_1,\ldots,x_m)$ of non-decreasing non-negative integers such that $x_m=n$, represented as a diagram of boxes. 
See Figure~\ref{boxandmatch} for better visual description and understanding. 
A {\it boxed diagram} can be defined as a pair of sequences $((x_1,\ldots,x_m),(y_1,\ldots,y_m))$ where $(x_1,\ldots,x_m)$ is a diagram and for all $i$ (such that $1\leq i \leq m$), the $y_i$-th box of the column $i$ of the diagram is marked. Note that $1 \leq y_i \leq x_i$.
Thus, a diagram can lead to $\prod_{i=1}^{m} x_i$ boxed diagrams. 
A {\it k-Dyck diagram} of size $n$ is defined as a diagram of width 
$m := (k-1)n+1$ and height $n$ such that $x_i \geq $ $\ceil{i/(k-1)}$ for all $i \leq m-1$. 
Finally, a {\it k-Dyck boxed diagram} of size $n$ is boxed diagram where the first coordinate $(x_1,\ldots,x_{(k-1)n+1})$ is a {\it k}-Dyck diagram of size $n$. Given these definitions, Bassino and Nicaud~\cite{BassinoN07} proved the following theorem.

\begin{theorem}~\cite{BassinoN07}
\label{bijection}
The set $\mathcal{D}_n$ containing DFAs having $n$ states and working over a $\sigma$-letter alphabet is in bijection with the set $\mathcal{B}_n$ of $\sigma$-Dyck boxed diagrams of size $n$. Moreover, the construction involving going from transition diagram of the DFA to $k$-Dyck boxed diagram and vice versa runs in linear time and space. 
\end{theorem}

\begin{figure}[htp]
	\begin{center}
		\includegraphics[scale=0.3]{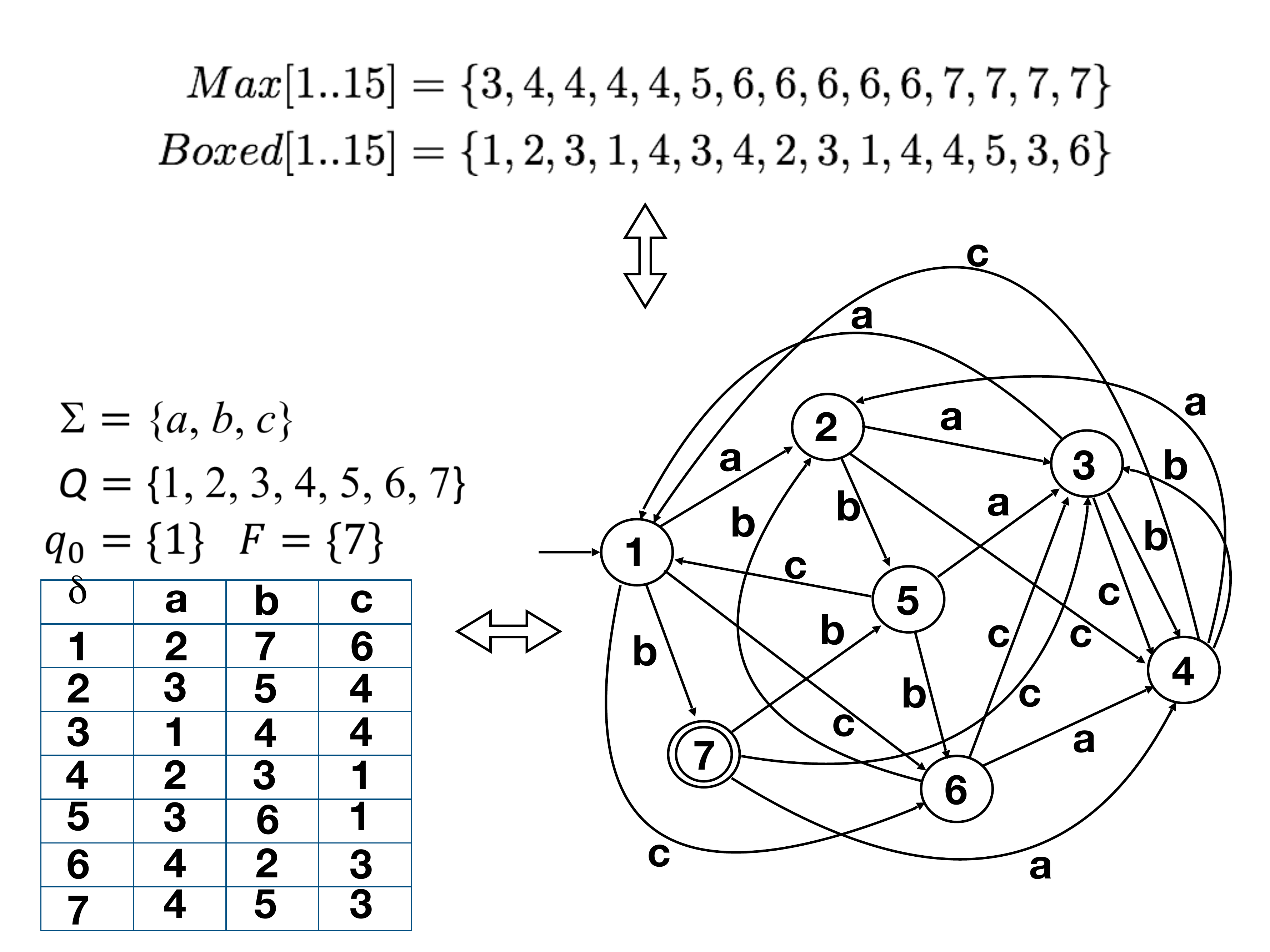}
	\end{center}
	\caption{Three ways to define the same DFA.  This DFA will serve as the working example for our discussion. By using the techniques of~\cite{BassinoN07}, this DFA can be entirely represented by the $\MAX[1..15]=\{3,4,4,4,4,5,6,6,6,6,6,7,7,7,7\}$ and $\BOXED[1..15]=\{1,2,3,1,4,3,4,2,3,1,4,4,5,3,6\}$ arrays of length $(\sigma-1) n +1=15$ each.}
	\label{fig_DFA1}
\end{figure}

Thus, by applying the above theorem, from any given DFA with $n$ states and $\sigma$-letter alphabet, \cite{BassinoN07} produces a $\sigma$-Dyck boxed diagrams of size $n$, which can be in turn  represented by two integer arrays $\MAX[1..m]$ and $\BOXED[1..m]$ of length $m := (\sigma-1) n +1$ each. Furthermore, from these two arrays, it is possible to entirely reconstruct the DFA using the algorithm of Theorem~\ref{bijection}. Thus, it is sufficient to store just these two arrays in order to encode any given DFA. For more details, readers are referred to~\cite{BassinoN07}. For an example, see Figure~\ref{fig_DFA1} which will also serve as the working example for this part of our paper. In particular, the DFA of Figure~\ref{fig_DFA1} can be entirely encoded by the $\MAX[1..15]=\{3,4,4,4,4,5,6,6,6,6,6,7,7,7,7\}$ and $\BOXED[1..15]=\{1,2,3,1,4,3,4,2,3,1,4,4,5,3,6\}$ arrays of length $(\sigma-1) n +1=15$, and these can be computed using the algorithms of~\cite{BassinoN07}.

First, we observe that, by construction, the arrays satisfy $1 \le \MAX[1] \le \MAX[2] \le \cdots \le \MAX[m] \le n$ and $1 \le \BOXED[i] \le \MAX[i]$ for each $i=1,2,\ldots,m$. This happens precisely because the translation is obtained by following a DFS on the DFA using the lexicographic order of words, and on each backtracking edge adding to the first vector the number of states scanned so far, and to the second vector the state reached. This also explains why each entry of these two arrays are upper bounded by $n$, the number of states of the given DFA. Now we consider the number of bits needed to encode the array $\MAX[1..m]$. As it is an increasing integer sequence of length $m$ and the range of the values is $[1,n]$, by using data structure $D(n, m, \epsilon)$ of Theorem~\ref{increasing}, this array can be represented using $O(\frac{1}{\epsilon}m^\epsilon n^{1-\epsilon}) = O(\frac{1}{\epsilon}\{(\sigma-1)n+1\}^\epsilon n^{1-\epsilon})$ bits of space. By letting $\epsilon = 1/\log(\sigma-1)$, the size is $O(n \log\sigma)$ bits if $\sigma>2$.
If $\sigma=2$, the space is obviously $O(n) = O(n \log\sigma)$ bits.
%, which is $o(n \log n)$ if $\log \sigma = o(\log n)$. 
Next we consider the number of bits required for array $\BOXED[1..m]$. Because each entry of this array is an integer from $1$ to $n$, we can use Theorem~\ref{vector_representation} to represent the $\BOXED[1..m]$ array using 
%$(\sigma-1)n \log n + O(\log^2 \sigma)$ 
$(\sigma-1)n \log n + O(\log^2 m)$ 
(recall $m = (\sigma-1) n +1$) bits.
%encode it in $\lceil \log n\rceil \le 1+\log n$ bits. The size of the array is therefore $m \lceil \log n\rceil \le ((\sigma-1)n+1)(1+\log n) = (\sigma-1)n (\log n + O(1))$ bits. 
Thus, in total, the size of the representation using two integer arrays is $(\sigma-1) n\log n + O(n \log \sigma)$
%$(\sigma-1)n (\log n + O(1)) + O(n \log\sigma)$ 
bits. Because the information theoretic lower bound is $(\sigma-1)n \log n + O(n)$ bits for the representation of DFA, this representation is succinct.
%if $\log \sigma = o(\log n)$.

We consider a special case when there is a failure/exit state labeled $0$ and
only $N$ transitions among all the $\sigma n$ transitions go to non-failure states.
Note that $\BOXED$ has $N-n+1$ non-zero values.
In this case we can reduce the space for $\BOXED[1..m]$ by using a new bitvector $Z[1..m]$
which has $N-n+1$ ones.  We use a new array $\BOXED'[1..N-n+1]$ which stores non-zero values of $\BOXED[1..m]$.
Then $\BOXED[i]$ is computed as follows.
If $Z[i]=0$, $\BOXED[i]=0$ (transition to the failure state).
If $Z[i]=1$, $\BOXED[i]=\BOXED'[partial\_rank_1(Z, i)]$.
If we use the data structure of Theorem~\ref{staticrs},
$Z$ is represented in $\sigma n+o(\sigma n)$ bits,
which is asymptotically smaller than the space lower bound of
$(\sigma-1)n \log n + O(n)$.
But, by using the data structure of Theorem~\ref{thm:id},  the bitvector $Z$ can be represented in
%sada $\log{\sigma n \choose N} + o(N) + O(\log\log (\sigma n))$ bits
$\log{\sigma n \choose N} + o(N) + O(\log\log (\sigma n)) = N \log \frac{\sigma n}{N} + O(N)$ bits
to support $partial\_rank$ queries in $O(1)$ time.
%
%By using the data structure $D(N-n+1, m, \epsilon)$, the bitvector $Z$ can be represented in
%$O(1/\epsilon \cdot (\sigma-1)^\epsilon n^\epsilon (N-n+1)^{1-\epsilon})$ bits
%and $rank_1(Z,i)$ is computed in constant time.
%By setting $\epsilon=1/\log(\sigma-1)$, the space is $O(N\log\sigma)$ bits.
%sada The space for $\BOXED'$ is $N \log n+O(N)$ bits.
The space for $\BOXED'$ is $(N-n+1) \log n$ bits.
Therefore the total space for representing a DFA with $N$ non-failure transitions is
%$N \log n +  \min\{\sigma n, O(N \log\sigma)\}$ bits.
%sada $N \log n + \log{\sigma n \choose N} + o(N) = N \log(\sigma n^2/N) + o(N)$ bits.
$(N-n) \log n + O(N \log\sigma)$ bits.

Even though this representation is optimal from the point of view of space occupancy, one major drawback of this representation is that, given a string $x$ over $\Sigma$, it takes linear time (in the size of the DFA, i.e., $O(\sigma n)$ time where $n$ is number of states of the DFA and $\sigma n$ is total number of transitions or edges in state transition diagram of the DFA) to decide whether the DFA accepts the string $x$, which is clearly not optimal as ideally it should be performed in time $O(|x|)$. This happens because the algorithm of Theorem~\ref{bijection} actually unravels the DFA from these two arrays $\MAX[1..m]$ and $\BOXED[1..m]$, and then checks whether the input string can be accepted or not. Thus, from the point of view of string acceptance, this encoding of DFA is not optimal whereas space requirement point of view, this is optimal. This motivates the need of a succinct encoding of a given DFA, where the problem of string acceptance can be performed in almost optimal time (i.e., almost in time proportional to the string length). In what follows, we provide such an encoding.

\subsection{Succinct Data Structure for DFA}\label{dfa_succinct}
{\bf Data structure:} To design a succinct data structure for DFA, we need the following three bitvectors $F$, $P$ and $T$ in addition to an integer array $\NBOXED[1..m]$ (that can be obtained from the $\BOXED[1..m]$ array of the previous section, as described later), which are defined as follows. 

$P$ is a balanced parentheses sequence of length $2n$ 
%(as defined in Theorem~\ref{succ_tree}) 
obtained from the lexicographic depth-first search (DFS) tree of the given input automaton $\mathcal{D}$. More specifically, given any DFA $\mathcal{D}$, we first perform the lexicographic DFS on $\mathcal{D}$ 
to generate the lexicographic DFS tree $R$ of $\mathcal{D}$,
i.e., while looking for a new edge to traverse during DFS, the algorithm always searches in lexicographic order of edge labels. For example, in Figure~\ref{fig_DFA1}, from any vertex, lexicographic DFS first tries to traverse the edge labeled $a$, followed by $b$ and finally $c$. 
The tree $R$ is represented as a balanced parenthesis sequence $P$ together with auxiliary structures to support the navigational queries on $R$, as mentioned in Theorem~\ref{succ_tree}, using $2n+o(n)$ bits.
%Then we can create the balanced parentheses sequence $P$ of $R$ in the usual way. 
The bitvector $F$ is used to mark all the final states of the input DFA, hence it takes $n$ bits.

\begin{figure}[tp]
	\begin{center}
		\includegraphics[scale=0.28]{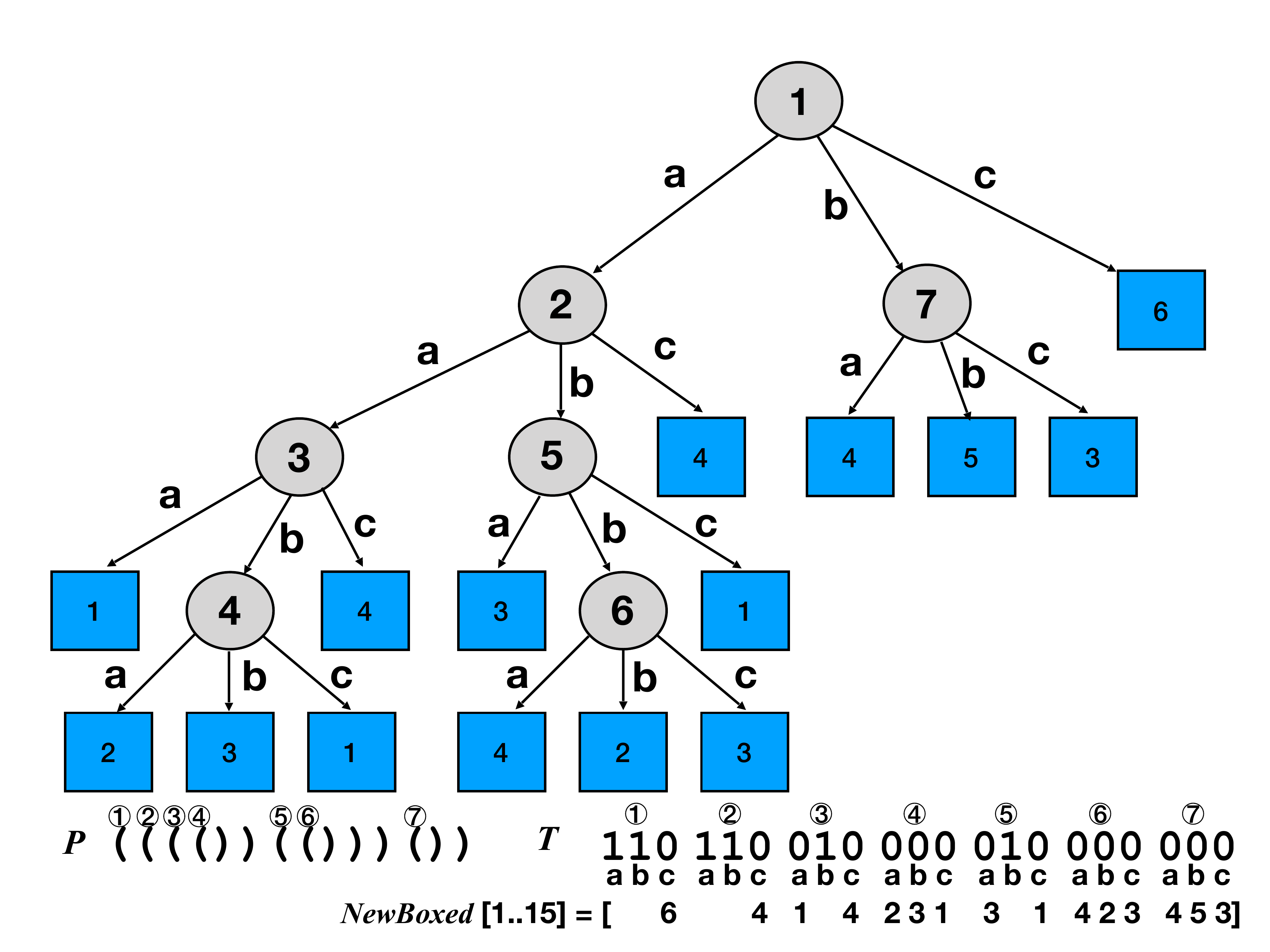}
	\end{center}
	\caption{The extended lex-DFS tree $S$ of the automaton of Figure~\ref{fig_DFA1} along with the corresponding bitvectors $P$, $T$, and the $\NBOXED[1..15]$ array (the elements of this array are drawn exactly below the corresponding $0$s with which they share one to one correspondence with). Note that, for the same automaton $\BOXED[1..15]$ array is given as $\BOXED[1..15]=\{1,2,3,1,4,3,4,2,3,1,4,4,5,3,6\}$.}
	\label{fig_DFA2}
\end{figure}

Before explaining the other bitvector, $T$, required for our succinct encoding, we want to explain the contents of Figure~\ref{fig_DFA2}. The tree depicted in the figure is what we call an {\it extended lexicographic DFS tree} or {\it extended lex-DFS tree} (denoted by $S$) in short. If we delete the squared nodes and their incident edges (originating from the circled nodes), we obtain the lexicographic DFS tree of the automaton $\mathcal{D}$. Actually these edges represent the {\it back edges/cross edges/forward edges}~\cite{CLRS} (i.e., non-tree edges) in the DFS tree of the automaton $\mathcal{D}$. Traditionally the vertices in the square are not drawn (as in our case of Figure~\ref{fig_DFA2}), rather the edges point to the nodes in the circle only (hence all the nodes appear only once). We have chosen to draw and define the extended lex-DFS tree this way as it helps us to design and explain our succinct data structure well. Also note that, edges originating from a circled node and going to another circled node represents tree edges whereas edges from circled to squared nodes represent non-tree %(i.e., back edges/cross edges/forward edges) 
edges. 

Now given the extended lex-DFS tree $S$, we visit the nodes of $S$ in DFS order and append a bit string of length $\sigma$ for each vertex $v$ of $S$ marking which of its children are attached to $v$ via tree edges (marked with $1$) and which are attached to $v$ via non-tree edges (marked with $0$) in the lexicographic order of the edge labels. The string obtained this way is referred to as $T$. Thus, $T$ is a bit-vector of length $\sigma n$ which captures the information about the tree and non-tree edges of $S$. More specifically, it has exactly $n-1$ ones, which have one-to-one correspondence with the tree edges of the lexicographic DFS tree of DFA $\mathcal{D}$, and has exactly $(\sigma-1) n +1$ zeros, which correspond to non-tree edges of the lexicographic DFS tree of DFA $\mathcal{D}$. See Figure~\ref{fig_DFA2} for an example. 
We relabel all the states of $\mathcal{D}$ such that the $i$-th vertex (state) in $R$ in preorder has label $i$, and also modify the transition function accordingly.
%We can reassign labels to the states/nodes (as they essentially are same object in our discussion) of the lexicographic DFS tree of DFA $\mathcal{D}$ by using the preorder numbers. 
Now it is easy to see that, for the state with label $i$ ($1 \le i \le n$), the corresponding node in the lexicographic DFS tree has exactly $\sigma$ outgoing edges, and we encode the tree edges among them using the bits in the range $T[\sigma (i-1)+1..\sigma i]$. More specifically, $T[\sigma (i-1) + c]=1$ if and only if the outgoing edge labeled $c$ is a tree edge ($1 \le c \le \sigma$). Similarly, we can also find the $j$-th outgoing tree edge from the state $i$ by $select_1(T,j + rank_1(T, \sigma (i-1)))$. 
Finally, we compress $T$ by observing that the positions of $1$s in the $T$ array form an increasing sequence, 
hence by using the data structure $D(n-1, \sigma n, \epsilon)$ of Theorem~\ref{increasing}, 
$access$, $rank$ and $select$ operations can be supported in constant time.
By setting $\epsilon = 1/\log (\sigma-1)$, $T$ can be encoded in $O(n \log\sigma)$ bits.
%which is the same size as $\MAX[\cdot]$.
%
%Finally, we compress $T$ by observing that the positions of $1$s in the $T$ array form an increasing sequence, hence by using Theorem~\ref{increasing}, $T$ can be encoded in $O(n \log\sigma)$ bits along with supporting $access$, $rank$ and $select$ operations in constant time in $T$. 
%
%{\bf SADA, could you please verify this compressing $T$ part for the correctness? I think we should put the explicit value of $\epsilon$ as used in Theorem~\ref{increasing} to get this bound as well. Also could you please add rank/select to the statement of  Theorem~\ref{increasing} as we need them in what follows. I was not very sure how to get them, hence asking you.}

Now let us define the new integer array $\NBOXED[1..m]$. First, observe 
%(it's implicit in~\cite{BassinoN07}) 
that elements of the array $\BOXED[1..m]$ are nothing but the leaves (i.e., node labels in the squared nodes) of the extended lex-DFS tree $S$ in the left to right order. More specifically, they are the node labels of the destinations of the non-tree edges emanating from the nodes of the lexicographic DFS tree of the automaton $\mathcal{D}$ in their preorder. %of their discovery during the lexicographic DFS. 
Instead of this specific ordering (followed in the $\BOXED[1..m]$ array), $\NBOXED[1..m]$ lists the same node labels in the order of their appearance in the $T$ bitvector (from left to right). Note that, as mentioned previously, these node are marked by $0$s in $T$ and they are in one-to-one correspondence with all the non-tree edges of the lexicographic DFS tree of the automaton $\mathcal{D}$. Thus, the $\NBOXED[1..m]$ array contains the same node labels as the $\BOXED[1..m]$ array, but in a different order. See Figure~\ref{fig_DFA2} for an example. This completes the description of our succinct data structure for DFA.
Note that $\MAX$ is no longer used in our data structure.

We now analyze the space complexity of our  
data structure. The array $\NBOXED[1..m]$ takes 
%$(\sigma-1)n \log n + O(\log^2 \sigma)$
$(\sigma-1)n \log n + O(\log^2 m)$
%$(\sigma-1)n (\log n + O(1))$ 
bits (by similar analysis as before for the $\BOXED[1..m]$ array). As mentioned previously, we store $T$ using Theorem~\ref{increasing}, hence it takes $O(n \log\sigma)$ bits.
The bitvector $F$ consumes $n$ bits. Finally, the bitvector $P$ is stored using Theorem~\ref{succ_tree}, hence it occupies $2n+o(n)$ bits in total. Thus, overall our data structure uses $(\sigma-1) n\log n + O(n \log \sigma)$
%$(\sigma-1)n (\log n + O(1))+O(n \log\sigma)+3n+o(n)$ 
bits. Hence, the data structure is succinct. 
It is easy to further reduce the size if the DFA has only $N < \sigma n$ non-failure transitions.
Using the bitvector $Z[1..m]$ for indicating non-failure transitions,
the array $\NBOXED[1..m]$ is compressed to 
%$N \log N$, 
$N-n+1$ non-zero values,
and the total space is
%sada $N \log n + \log{\sigma n \choose N} + o(N) = N \log(\sigma n^2/N) + o(N)$ bits.
$(N-n) \log n + O(N \log\sigma)$ bits.
%$N \log n +  \min\{O(\sigma n), O(N \log\sigma)\}$ bits.
In what follows, we describe the string acceptance query algorithm using our data structures.
%$S(n,\sigma n)$, 
\\

\noindent
{\bf Query algorithm.} Suppose we are given an input string $x$ of length $y$ over $\Sigma$, and we need to decide if the DFA $\mathcal{D}$ accepts $x$ or not. We start the following procedure from the initial state
(stored explicitly using $O(\log n)$ bits) and repeat until the end of the input string $x$. At any generic step, to figure out the transition function $\delta(q, c) := q'$ where $1 \le q, q' \le n$ are 
the states, we first look at the bit $T[\sigma (q-1) + c]$. If it is $1$, the outgoing edge labeled $c$ from state $q$ is a tree edge. Let $j := rank_1(T, \sigma (q-1) + c) - rank_1(T, \sigma (q-1))$. Then the outgoing edge is the $j$-th tree edge of node $q$ in the lex DFS tree.
Therefore $q' = child(q, j)$ (supported using the Theorem~\ref{succ_tree}). If the bit is $0$, the outgoing edge labeled $c$ from state $q$ is a non-tree edge. Let $j := rank_0(T, \sigma (q-1) + c)$.  Then the edge is the $j$-th non-tree edge
in the DFA, and $q'$ is obtained by 
%$q' := \BOXED[j]$. 
$q' := \NBOXED[j]$. 
%All of this can be done in constant time. 
Hence, when we reach the end of $x$, and if we are at an accepting/final states (can be figured out from the bitvector $F$), we say that the DFA $\mathcal{D}$ accepts $x$. 
The $rank$ operations on $T$ take $O(\log \sigma)$ time while all other operations, at each step, take $O(1)$ time.
Thus the overall run time for checking the membership of an input string $x$ is $O(|x| \log \sigma)$. This completes the proof of Theorem~\ref{dfa_main_result}.

%We summarize our result in the following theorem.
%\begin{theorem}
%Given a non-isomorphic, initially-connected deterministic finite automata $\mathcal{D}$ having $n$ states and working over an alphabet $\Sigma$ of size $\sigma$, there exists a succinct encoding for $\mathcal{D}$ taking $(\sigma-1)n (\log n + O(1))+O(n \log\sigma)+3n+o(n)$ bits of space, which can determine, given an input string $x$ over $\Sigma$, whether $\mathcal{D}$ accepts $x$ optimally in time proportional to the length of $x$ along with using constant words of working space.
%\end{theorem}

{\bf Remark:} In the light of the above discussion, consider the following. Suppose we are given as input a succinct representation for a DFA $\mathcal{D}$ whose language is $\mathcal{L}(\mathcal{D})$, and our goal is to construct the succinct representation for the DFA (say $\mathcal{D}'$) which accepts complement of $\mathcal{L}(\mathcal{D})$ i.e., $\mathcal{L}(\mathcal{D}')= \Sigma^*-\mathcal{L}(\mathcal{D})$. In order to construct the succinct representation for $\mathcal{D}'$, we start with the succinct representation for $\mathcal{D}$ (that is given in terms of three bit vectors $F, P, T$ and the integer array $\NBOXED[1..m]$), and simply convert (in the $F$ array) each final state in $\mathcal{D}$ into a non-final state in $\mathcal{D}'$ and convert each non-final state in $\mathcal{D}$ into a final state in $\mathcal{D}'$ without changing any other data structures. As a consequence, it is easy to see that, we will end up with what we desired.

\subsection{Succinct Data Structures for Acyclic DFA}\label{acyclic_dfa_succinct}
As mentioned previously, an acyclic DFA $\mathcal{A}$ with total $n$ states always has a unique dead state and $n-1$ transient (i.e., non dead) states. Another way to visualize $\mathcal{A}$ is to see that the state transition diagram of $\mathcal{A}$ does not have any cycles except at the unique dead state. Given such a setting, one can always use the succinct encoding (of the previous section) of an arbitrary DFA 
%(possibly containing cycles in the state transition diagram) 
to represent them. In that case, we end up using $(\sigma-1) n\log n + O(n \log \sigma)$ bits of space. In what follows, we show that by exploiting the acyclic property, one can obtain improved space bound for representing $\mathcal{A}$.

We basically view the state transition diagram of $\mathcal{A}$ as a directed acyclic graph with a single source (i.e., the initial state), and a single sink i.e., the dead state (call it $d$). Given this, we first construct a spanning tree $W = (V, E)$ of $\mathcal{A}$ where $V = Q$ (i.e., the set of states of $\mathcal{A}$) and $E = \{ (q_u, q_v) \mid \delta(q_v, \sigma) = q_u \mbox{ where } q_v \neq d\}$ by making the dead state $d$ as the root of this tree. It is easy to see that such a spanning tree can always be constructed. By applying Theorem~\ref{succ_tree}, we encode the structure of $W$ using $2n+o(n)$ bits to support the navigational queries on $W$ (in particular, the parent query) in $O(1)$ time. As done previously in Section~\ref{dfa_succinct} while constructing the succinct data structures for DFA, here also we relabel all the states of $\mathcal{A}$ such that the $i$-th vertex (state) in $W$ in preorder has label $i$, and modify the transition function accordingly. Note that the dead state $d$ is labeled with label $0$ in this ordering, and we do not need to store the transition function for the dead state. We also mark in a bitvector of size $n$ all the final states of $\mathcal{A}$, and 
%use also %sada
we store the label of the start state. We then store a two dimensional array $L[1..n-1][1..\sigma-1]$ such that $L[q][i] = \delta(q, i)$ using data structure of Theorem~\ref{vector_representation}. Thus, the overall space usage is
$(\sigma -1)(n-1) \log n + 3n +O(\log^2 \sigma)+ o(n)$ bits. 

In what follows, we explain how to check if $\mathcal{A}$ accepts any given string $x$ over $\Sigma$. At any generic step, to compute $\delta(q,i)$, we simply output $L[q][i]$ if $i \in  \{1, 2, \dots, \sigma-1 \}$; otherwise (i.e., if $i=\sigma$) the value of $\delta(q, \sigma)$ is given by the parent of $q$ in $W$ i.e., $\delta(q, i)=parent(q)$. Thus $\delta(q,i)$ can be computed in constant time, and hence we can optimally decide if $\mathcal{A}$ accepts $x$ in time proportional to the length of $x$. This completes the proof of Theorem~\ref{acyclic_dfa_main_result}. 

% lower bound ..??

%Let $\mathcal{A}$ be the given acyclic DFA that we want to encode, with start state $s$; and let $\Sigma = \{c_1, c_2, \dots, c_\sigma \}$ be the alphabet.  Assume that there is a single ``dead state'', $d$ (if there are multiple dead states, we can `merge' all those dead states into a single dead state). We first construct a spanning tree $T = (V, E)$ of $\mathcal{A}$ where $V = Q$ and $E = \{ (u, v) | \delta(v, c_\sigma) = u \mbox{ and } v \neq d\}$. The dead state $d$ is the root of this tree. We encode the structure of this tree using $2n+o(n)$ bits to support the navigational queries on the tree (in particular, the parent query) in $O(1)$ time. We relabel all the states in $Q$ such that the $i$-th vertex (state) in $T$ in preorder has label $i$ (and modify the transition function accordingly). We then store a two dimensional array $A[1..n][1..\sigma-1]$ such that $A[q][i] = \delta(q, c_i)$. Thus the overall space usage is $(\sigma -1)n \log n + 2n + o(n)$ bits.

%To compute $\delta(q,c)$, we simply output $A[q][c]$ if $c \in  \{c_1, c_2, \dots, c_{\sigma-1} \}$; otherwise $\delta(q, c_\sigma)$ is given by the parent of $q$ in $T$.

\subsection{Succinct Encoding for NFA}\label{nfa_succinct}
As mentioned previously in Section~\ref{enumeration}, to encode an initially connected NFA on $n$ states over a $\sigma$-letter alphabet $\Sigma$ with a fixed initial state and one or more final states, we need at least $\sigma n^2+n$ bits. In what follows, we show a very simple scheme achieving this bound. 

We store a table $H$ having $n$ rows (corresponding to the $n$ states of the input NFA) and $\sigma$ columns (corresponding to each letter of the alphabet $\Sigma$). The entry $H[i][j]$ (where $0 \leq i \leq n-1$ and $1 \leq j \leq \sigma$) basically stores the corresponding transition function of the NFA i.e., $H[i][j]=\delta(q_i,j)$ where $q_i \in Q$ and $j \in \Sigma$. Now for an NFA, 
$\delta(i,j)$ is a subset of $Q$. If we store this subset explicitly, it might take $O(n \log n)$ bits in the worst case per transition of the NFA, leading to overall $\sigma n^2 \log n$ bits which is $O(\log n)$ multiplicative factor off from the optimal space requirement. Instead we simply store the charecteristic vector $L$ of the subset (of length $n$, marking the corresponding states from the subset as $1$, and rest of the bits in $L$ are $0$) where the state labeled $i$ of the NFA moves to after reading the letter $j \in \Sigma$. Thus, the overall size of $H$ is exactly $\sigma n^2$ bits. Finally, we also mark in a separate bitvector (of length $n$) all the final states of the input NFA. Thus, in total the size of our encoding is given by $\sigma n^2+n$ bits, which matches the lower bound. Hence, our encoding is succinct and optimal. 

Now using our encoding, we can simply implement the classical algorithm (given in the texts of~\cite{Hopcroft,Sipser}) for checking if the NFA accepts a given input string or not, and this runs in $O(n^2 |x|)$ time where $x$ is the input string and $|x|$ denotes its length. Note that we also need two bitvectors of length $n$ each (hence overall $2n$ bits) as working space to mark two sets of intermediate states between successive transitions while executing the string acceptance checking algorithm. Hence, we obtain the result mentioned in Theorem~\ref{nfa_main_result}.

\section{Concluding Remarks}\label{conclusion}
We considered the problem of succinctly encoding any given DFA $\mathcal{D}$, acyclic DFA $\mathcal{A}$ or NFA $\mathcal{N}$ so as to check efficiently if they accept a given input string. To this end, we successfully designed succinct data structures for them that also support the string acceptance query efficiently for DFAs, acyclic DFAs,
% $\mathcal{A}$, 
and NFAs.
%$\mathcal{N}$.
%, matching the running times of the classical algorithms. 
To the best of our knowledge, our work is the first attempt to encode any mathematical models from the world of automata theory using the lens of succinct data structures, and we believe that our work will spur further interest in other similar problems in future.

\bibliography{dfs}

\begin{thebibliography}{10}

\bibitem{Acan}
H.~Acan, S.~Chakraborty, S.~Jo, and S.~R. Satti.
\newblock Succinct data structures for families of interval graphs.
\newblock In {\em WADS}, 2019.

\bibitem{AleardiDS08}
L.~C. Aleardi, O.~Devillers, and G.~Schaeffer.
\newblock Succinct representations of planar maps.
\newblock {\em Theor. Comput. Sci.}, 408(2-3):174--187, 2008.

\bibitem{BassinoN07}
F.~Bassino and C.~Nicaud.
\newblock Enumeration and random generation of accessible automata.
\newblock {\em Theor. Comput. Sci.}, 381(1-3):86--104, 2007.

\bibitem{Clark96}
D.~R. Clark.
\newblock {\em Compact Pat Trees}.
\newblock PhD thesis. University of Waterloo, Canada, 1996.

\bibitem{CLRS}
T.~H. Cormen, C.~E. Leiserson, R.~L. Rivest, and C.~Stein.
\newblock {\em Introduction to Algorithms {(3.} ed.)}.
\newblock {MIT} Press, 2009.

\bibitem{Diestel}
R.~Diestel.
\newblock {\em Graph Theory, 4th Edition}, volume 173 of {\em Graduate texts in
  mathematics}.
\newblock Springer, 2012.

\bibitem{DietzfelbingerKMHRT94}
M.~Dietzfelbinger, A.~R. Karlin, K.~Mehlhorn, F.~{Meyer auf der Heide},
  H.~Rohnert, and R.~E. Tarjan.
\newblock Dynamic perfect hashing: Upper and lower bounds.
\newblock {\em {SIAM} J. Comput.}, 23(4):738--761, 1994.

\bibitem{DodisPT10}
Y.~Dodis, M.~Patrascu, and M.~Thorup.
\newblock Changing base without losing space.
\newblock In {\em STOC}, pages 593--602, 2010.

\bibitem{Domaratzki06}
M.~Domaratzki.
\newblock Enumeration of formal languages.
\newblock {\em Bulletin of the {EATCS}}, 89:117--133, 2006.

\bibitem{DomaratzkiKS02}
M.~Domaratzki, D.~Kisman, and J.~Shallit.
\newblock On the number of distinct languages accepted by finite automata with
  n states.
\newblock {\em Journal of Automata, Languages and Combinatorics},
  7(4):469--486, 2002.

\bibitem{FarzanK14}
A.~Farzan and S.~Kamali.
\newblock Compact navigation and distance oracles for graphs with small
  treewidth.
\newblock {\em Algorithmica}, 69(1):92--116, 2014.

\bibitem{FarzanM13}
A.~Farzan and J.~I. Munro.
\newblock Succinct encoding of arbitrary graphs.
\newblock {\em Theor. Comput. Sci.}, 513:38--52, 2013.

\bibitem{Flajolet}
P.~Flajolet and R.~Sedgewick.
\newblock {\em Analytic Combinatorics}.
\newblock Cambridge University Press, 2009.

\bibitem{Hopcroft}
J.~E. Hopcroft, R.~Motwani, and J.~D. Ullman.
\newblock {\em Introduction to automata theory, languages, and computation -
  international edition {(2.} ed)}.
\newblock Addison-Wesley, 2003.

\bibitem{Jacobson}
G.~J. Jacobson.
\newblock {\em Succinct static data structures}.
\newblock PhD thesis. Carnegie Mellon University, 1998.

\bibitem{Liskovets06}
V.~A. Liskovets.
\newblock Exact enumeration of acyclic deterministic automata.
\newblock {\em Discrete Applied Mathematics}, 154(3):537--551, 2006.

\bibitem{MunroRRR12}
J.~I. Munro, R.~Raman, V.~Raman, and S.~S. Rao.
\newblock Succinct representations of permutations and functions.
\newblock {\em Theor. Comput. Sci.}, 438:74--88, 2012.

\bibitem{MunroR01}
J.~I. Munro and V.~Raman.
\newblock Succinct representation of balanced parentheses and static trees.
\newblock {\em {SIAM} J. Comput.}, 31(3):762--776, 2001.

\bibitem{MunroW18}
J.~I. Munro and K.~Wu.
\newblock Succinct data structures for chordal graphs.
\newblock In {\em ISAAC}, pages 67:1--67:12, 2018.

\bibitem{Navarro}
G.~Navarro.
\newblock {\em Compact Data Structures - {A} Practical Approach}.
\newblock Cambridge University Press, 2016.

\bibitem{NavarroS14}
G.~Navarro and K.~Sadakane.
\newblock Fully functional static and dynamic succinct trees.
\newblock {\em {ACM} Transactions on Algorithms}, 10(3):16, 2014.

\bibitem{RamanRS07}
R.~Raman, V.~Raman, and S.~R. Satti.
\newblock Succinct indexable dictionaries with applications to encoding
  \emph{k}-ary trees, prefix sums and multisets.
\newblock {\em {ACM} Trans. Algorithms}, 3(4):43, 2007.

\bibitem{Sipser}
M.~Sipser.
\newblock {\em Introduction to the theory of computation}.
\newblock {PWS} Publishing Company, 1997.

\bibitem{Sumigawa18}
K.~Sumigawa and K.~Sadakane.
\newblock An efficient representation of partitions of integers.
\newblock In {\em IWOCA}, pages 361--373, 2018.

\end{thebibliography}
\newpage
\appendix
\section{Appendix} 
\label{appendix}

\subsection{Supporting More Operations (Union and Intersection)}\label{product_construction}
\begin{figure}[tp]
	\begin{center}
		\includegraphics[scale=0.35]{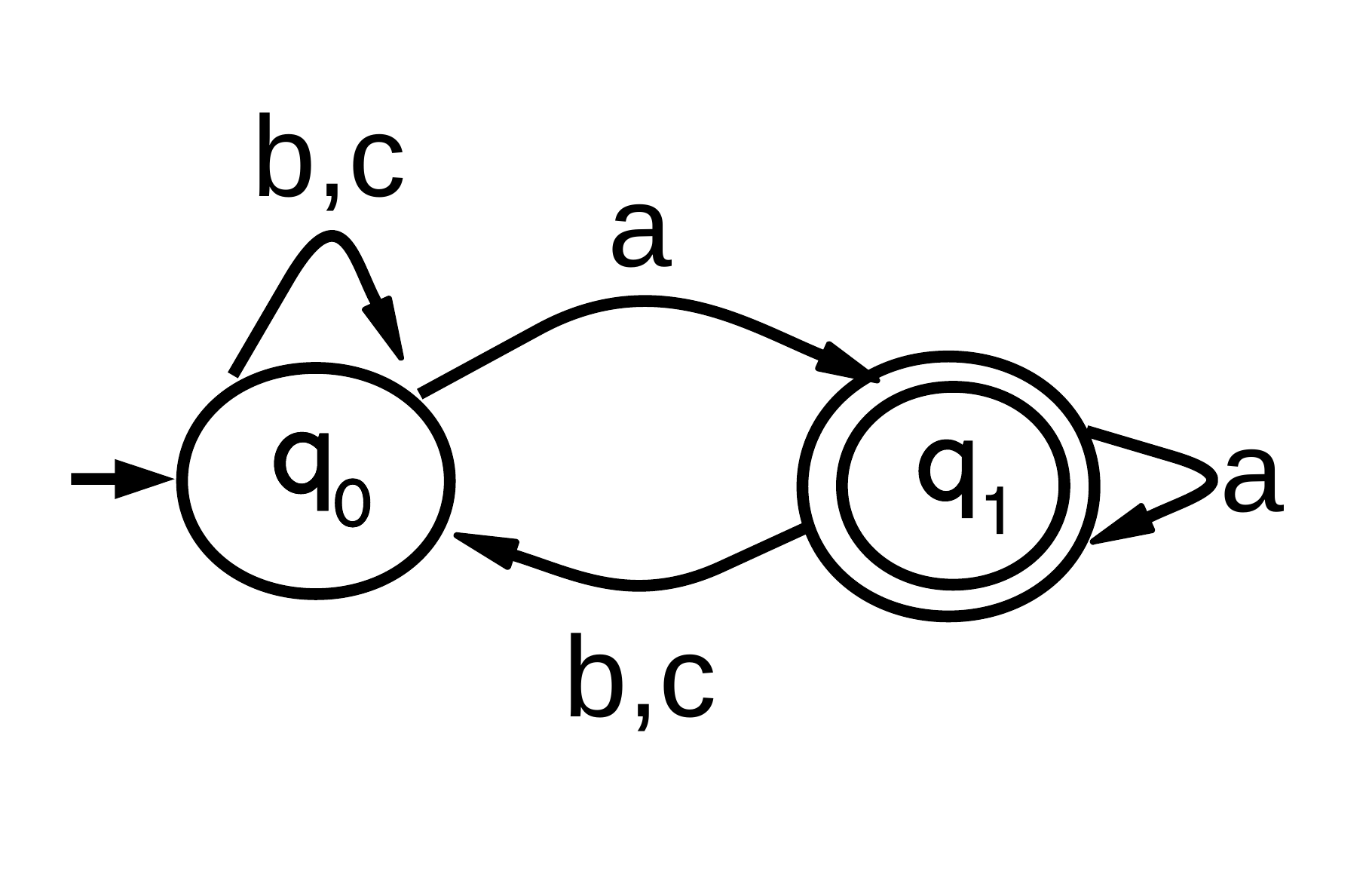}
		\includegraphics[scale=0.35]{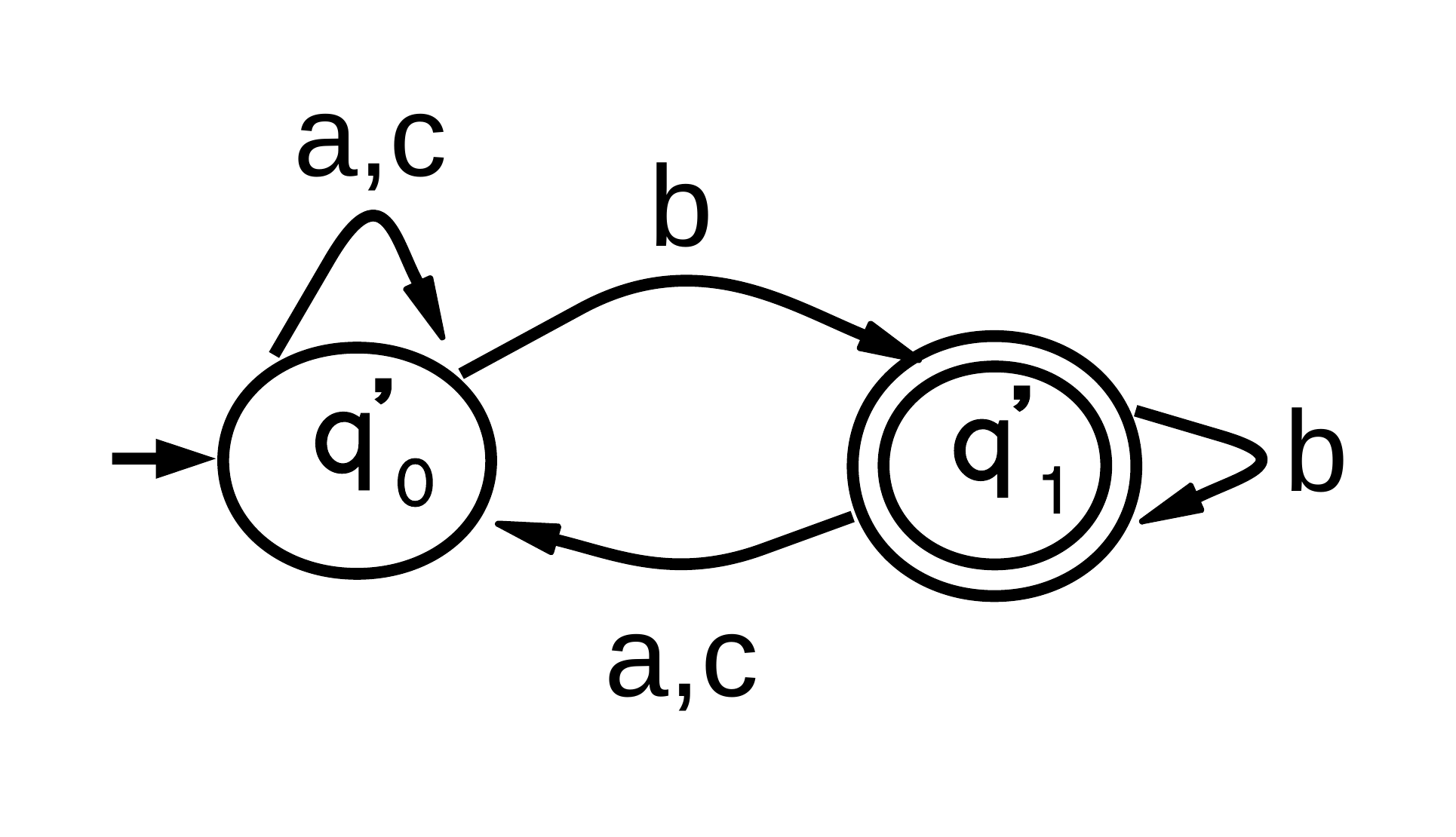}
	\end{center}
	\caption{The state transition diagram for a DFA $\mathcal{D}_1$ (on the left) where $\mathcal{D}_1 =(\Sigma, Q, q_0, \delta, F)$ such that (i) $\Sigma=\{a,b,c\}$, (ii) $Q=\{q_0,q_1\}$, (iii) $q_0=q_0$ (marked with an incoming arrow coming from nowhere), (iv) $F=\{q_1\}$, and (v) the transition function is defined as the following set, $\{\delta(q_0,a)=q_1,\delta(q_0,b)=q_0,\delta(q_0,c)=q_0, \delta(q_1,a)=q_1,\delta(q_1,b)=q_0,\delta(q_1,c)=q_0\}$. Precisely the DFA $\mathcal{D}_1$ accepts all the strings that end with an $a$ over $\Sigma$. Similarly the state transition diagram for a DFA $\mathcal{D}_2$ (on the right) where $\mathcal{D}_2 =(\Sigma, Q', q'_0, \delta', F')$ such that (i) $\Sigma=\{a,b,c\}$, (ii) $Q=\{q'_0,q'_1\}$, (iii) $q'_0=q'_0$ (marked with an incoming arrow coming from nowhere), (iv) $F=\{q'_1\}$, and (v) the transition function is defined as the following set, $\{\delta(q'_0,a)=q'_0,\delta(q'_0,b)=q'_1,\delta(q'_0,c)=q'_0, \delta(q'_1,a)=q'_0,\delta(q'_1,b)=q'_1,\delta(q'_1,c)=q'_0\}$. Precisely the DFA $\mathcal{D}_2$ accepts all the strings that end with a $b$ over $\Sigma$. These two DFAs will serve as the working example for our discussion of the product automaton construction.}
	\label{fig_DFA7}
\end{figure}

\begin{figure}[tp]
	\begin{center}
 	\includegraphics[scale=0.4]{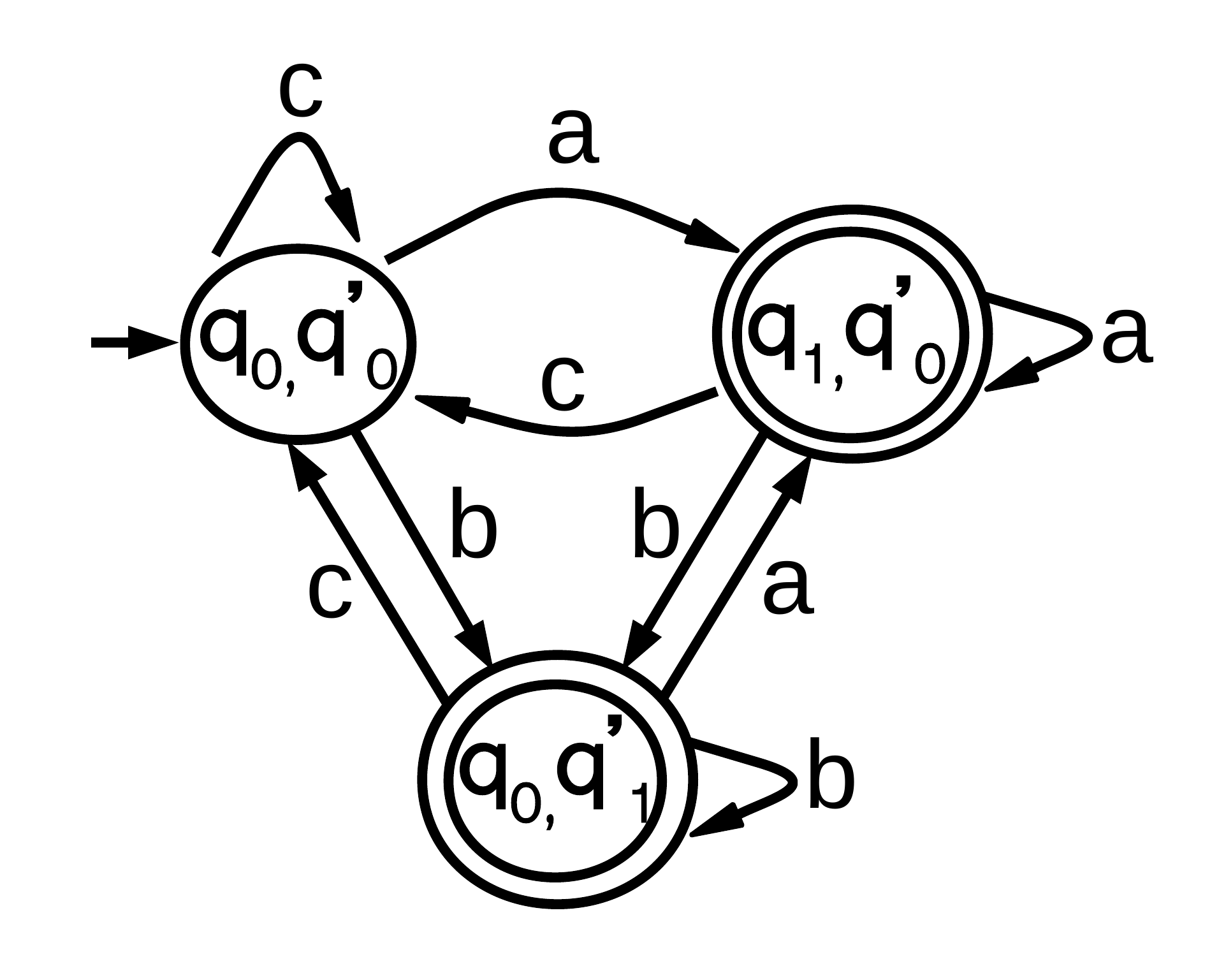}
	\includegraphics[scale=0.4]{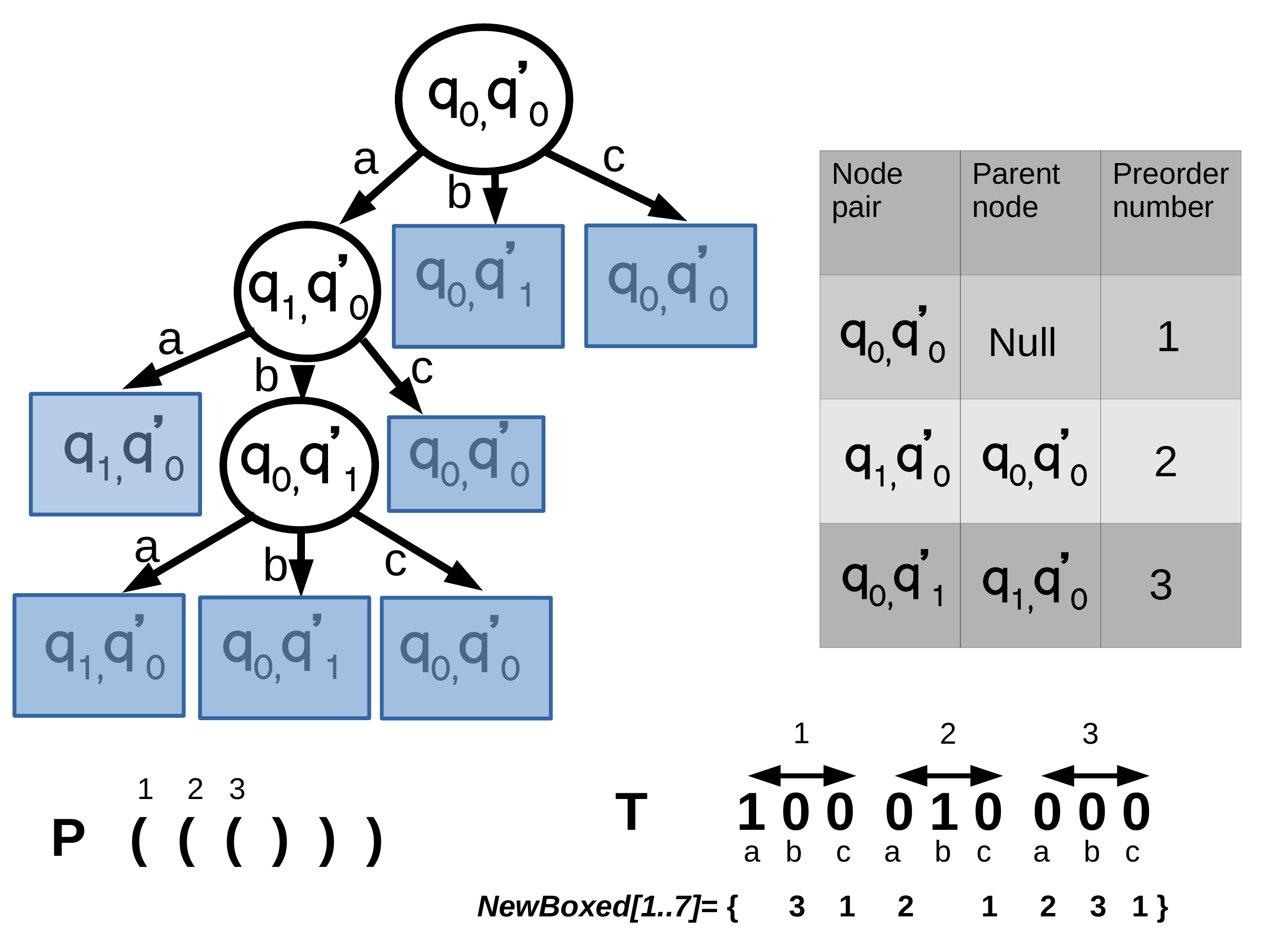}
	\end{center}
	\caption{The state transition diagram on the top depicts the product automaton $\mathcal{P}$ accepting the language $\mathcal{L}(\mathcal{D}_1) \cup \mathcal{L}(\mathcal{D}_2)$ (DFA $\mathcal{D}_1$ and $\mathcal{D}_2$ are defined in Figure~\ref{fig_DFA7}). The diagram on the bottom left depicts the extended lex-DFS tree of the product automaton $\mathcal{P}$ (defined above) whereas the rest of diagram contains the description of the other data structures i.e., the bitvectors $P,T$, the integer array $\NBOXED[1..7]$, and finally the hash table. Note that the elements of the $\NBOXED[1..7]$ array are drawn exactly below the corresponding $0$s with which they share one to one correspondence with.}
	\label{fig_DFA8}
\end{figure}

In what follows we show how to support some standard operations on DFAs space efficiently. We start with the classical example of {\it product automaton} construction. More specifically, given the succinct representation of two DFAs, we want to construct a succinct representation of the product automaton accepting the language which is the union/intersection of the two input DFA's language. Before providing our construction, let us formally define the product automaton construction. Suppose, we are given two DFAs $\mathcal{D}_1=(\Sigma, Q, q_0, \delta, F)$ and $\mathcal{D}_2=(\Sigma, Q', q'_0, \delta', F')$ represented succinctly (as described in Section~\ref{dfa_succinct}) and both working over the same alphabet $\Sigma$. Then a product automaton (denoted by $\mathcal{P}$) of $\mathcal{D}_1$ and $\mathcal{D}_2$ is defined as follows, $\mathcal{P}=(\Sigma, \mathcal{Q}, (q_0,q'_0), \delta_p, F_p)$ where $\mathcal{Q}= Q \times Q'$, and $\delta_p : \mathcal{Q} \times \Sigma \rightarrow \mathcal{Q}$. Moreover, for any $q \in Q, q' \in Q'$ and $c \in \Sigma$, $\delta_p ((q,q'), c) := (\delta(q,c), \delta'(q',c))$. The start state of $\mathcal{P}$ is the pair $(q_0,q'_0)$ whereas the final state can be defined in multiple ways. More specifically, if we set $F_p=F \times F'$, then $\mathcal{L}(\mathcal{P})= \mathcal{L}(\mathcal{D}_1) \cap \mathcal{L}(\mathcal{D}_2)$. Similarly, if we set $F_p= (F \times Q') \cup (Q \times F')$, then $\mathcal{L}(\mathcal{P})= \mathcal{L}(\mathcal{D}_1) \cup \mathcal{L}(\mathcal{D}_2)$. Now we show how one can directly construct a succinct representation of $\mathcal{P}$  given the succinct representations of $\mathcal{D}_1$ and $\mathcal{D}_2$ as input, and note that, to do so we just need to describe how one can create the three bitvectors $F, P, T$ and the integer array $\NBOXED[1..m]$ corresponding to $\mathcal{P}$ from the succinct representations of $\mathcal{D}_1$ and $\mathcal{D}_2$ directly. See Figure~\ref{fig_DFA7} and Figure~\ref{fig_DFA8} for a visual description of our product automaton construction algorithm.

For constructing the product automaton $\mathcal{P}$, our high level idea is to create the states and transitions of $\mathcal{P}$ by generating the states of $\mathcal{P}$ in the lexicographic DFS order using two passes. In the first pass, we generate the $P$ and $T$ arrays (both initialized with empty string), and this is followed by the construction of  the $\NBOXED[1..m]$ array in the second pass. More specifically, we start by creating the initial state i.e., $(q_0,q'_0)$ as the first circled node i.e., root in the extended lex-DFS tree corresponding to $\mathcal{P}$, store an entry corresponding to this node in the hash table along with storing its preorder number (which is $1$ in the case of $(q_0,q'_0)$) as a satellite data in the hash table. Also we append $\sigma$ zero bits to $T$ corresponding to the root. In general, at any point of time during the execution of this algorithm, the hash table stores an entry corresponding to each of the circled nodes generated upto that point along with storing its preorder number and its parent node as satellite data. Note that for the root, we don't need to store any parent information. Now to figure out the transitions out of any state, note that, if we use the method described in the query algorithm for DFA (as described in Section~\ref{dfa_succinct}) we need to pay $O(\log \sigma)$ time per symbol of the alphabet $\Sigma$. Instead, in what follows, we show how one can find each transition in $O(1)$ time per symbol out of any state using all the information that is already stored in the input i.e., succinct representations for $\mathcal{D}_1$ and $\mathcal{D}_2$. 
Assume for now that we can do so and also suppose that at some point of the algorithm, we created a new circled node $(i,i')$. Then we proceed as follows. First we append $\sigma$ zero bits to the bit string $T$ corresponding to the node $(i,i')$. This is followed by the {\em expansion} of the state $(i,i')$ by generating the transitions $\delta_p((i,i'), c)$ in the lexicographic ordering of the alphabet characters $c \in \Sigma$, as follows. Let $j=\delta (i,c)$ and $j'=\delta'(i',c)$, then we check in the hash table if the state $(j,j')$ has already been created before (by checking membership in the hash table). If yes, we create a squared node $(j,j')$ as a child node of $(i,i')$ (which is a circled node) and don't make any changes to the $P$ array, mark the $c$-th bit corresponding to the node $(i,i')$ in $T$ as zero; and continue with the expansion of $(i,i')$ with the next character in $\Sigma$. If not, we create a circled node $(j,j')$ as a child of $(i,i')$, append an open parenthesis to the $P$ array constructed so far, mark the $c$-th bit corresponding to the node $(i,i')$ in $T$ as one, and finally insert $(j,j')$ into the hash table along with inserting $(i,i')$ as its parent and its preorder number as its satellite data; and continue with the expansion of $(j,j')$. Finally, when we exhaust checking all the characters $c \in \Sigma$ out of $(i,i')$, we backtrack to the parent of $(i,i')$ in the extended lex-DFS tree (using the parent information stored as a satellite data with the entry for the node $(i,i')$), and in this case, we simply append a close parenthesis to the $P$ array constructed so far. It is clear that using this procedure repeatedly we can successfully create $P$ and $T$ arrays corresponding to the product automaton $\mathcal{P}$. Finally, we create the all the auxiliary structures (mentioned in Section~\ref{prelims_stricture}) on top of the arrays $P$ and $T$ (similar to the succinct data structure for DFA as described in Section~\ref{dfa_succinct}) for supporting various navigational queries on the extended lex-DFS tree. Intuitively the $P$ array stores the topology of the extended lex-DFS tree of the state transition diagram of the product automaton $\mathcal{P}$ and the $T$ array stores the parent-child relationship between the nodes of the extended lex-DFS tree in a compact manner. Now let's discuss how to find out the transitions efficiently. Note that it suffices to describe how one can find $j = \delta (i,c)$ in $\mathcal{D}_1$  ($j' = \delta' (i',c)$ in $\mathcal{D}_2$ can be found similarly). We consider the two cases: when the edge $(i,j)$ is a (i) non-tree edge, or a (ii) tree edge. 
In case (i), 
%assume $(i,j)$ is a non-tree edge in the extended lex-DFS tree corresponding to $\mathcal{D}_1$, then 
$j=\NBOXED[rank_0(T, \sigma (i-1) + c)]$. 
In case (ii), 
%assume $(i,j)$ is a tree edge in the extended lex-DFS tree corresponding to $\mathcal{D}_1$, 
$j= child(i, t)$ (can be supported using the Theorem~\ref{succ_tree} on the $P$ array) where $t = rank_1(T, \sigma (i-1) + c) - rank_1(T, \sigma (i-1))$.

In what follows, we describe how one can fill up the integer array $\NBOXED[1..m]$ with $m$ (we discuss about fixing $m$ later) entries which are initialized with all one. Note that, similar to the succinct DFA construction, this array should contain the preorder number of the node labels in the squared nodes of the extended lex-DFS tree in the order of their appearance in the $T$ bitvector (from left to right). Moreover, these node are marked by $0$s in $T$ and they are in one-to-one correspondence with all the non-tree edges of the extended lex-DFS tree of the product automaton $\mathcal{P}$. To fill up $\NBOXED[1..m]$ array, we follow essentially the same lexicographic DFS traversal procedure as we did in the first pass except the following. More specifically, we start the second pass of the extended lex-DFS tree and whenever we encounter a non-tree edge, we retrieve the preorder number corresponding to the node label in the squared node (i.e., the other end point of that non-tree edge) from the hash table, and insert this number at the suitable position in the $\NBOXED$ array. In detail, suppose we are at a circled node $(i,i')$ (with preorder number, say, $k$) and currently exploring the transition with the letter $c \in \Sigma$ out of $(i,i')$. Also assume that $\delta_p((i,i'), c)=(j,j')$ and $(j,j')$ is a squared node (i.e., $((i,i'),(j,j'))$ is a non-tree edge) such that the preorder number associated with the node label $(j,j')$ is $d$ in the hash table. Then, we assign $ \NBOXED[\ell]=d $ where $\ell = rank_0(T, \sigma (k-1) + c)$. Finally, depending on union or intersection operation, we also mark in another bitvector $F$ (according to the definition given above) all the final states of the product automaton $\mathcal{P}$. Observe that once we have all the constituent data structures (including all the auxiliary data structures that we build on top of $F, P, T$ arrays and the integer array $\NBOXED[1..m]$) for the succinct representation for $\mathcal{P}$ ready, we can essentially use the same query algorithm for string acceptance checking as we described for DFA in Section~\ref{dfa_succinct}.

Let's analyze the resource requirements for our algorithm. Suppose $|Q|=n$ and $|Q'|=n'$, then the product automaton $\mathcal{P}$ can have $nn'$ states at the worst case, but in general it could be much less as well. Let us suppose that $\mathcal{P}$ has  $n''$ states, then $n'' \leq nn'$, and in what follows, we write our space requirement as a function of $n''$. If we implement the hash table using the data structure of~\cite{DietzfelbingerKMHRT94}, then it consumes $O(n''\log n'')$ bits in total. Also note that this is the dominating term for the working space bound as other auxiliary data structures consume negligible space with respect to the space consumption for the hash table. Moreover, our algorithm runs in linear (in $n''$) expected time overall. The randomized nature of our algorithm is due to the fact of using the hashing data structure of~\cite{DietzfelbingerKMHRT94} whereas all the other parts of our algorithm is deterministic. As a result of our algorithm, we generate a representation for $\mathcal{P}$ and this is given by the following arrays. The bitvectors $P$ and $F$ consume $2n''+o(n'')$, $n''$ bits respectively. For the $T$ array, we compress it by observing that the positions of $1$s in the $T$ array form an increasing sequence, hence by using the data structure $D(n''-1, \sigma n'', \epsilon)$ of Theorem~\ref{increasing}, $access$, $rank$ and $select$ operations can be supported in constant time, and by setting $\epsilon = 1/\log (\sigma-1)$, $T$ can also be encoded in $O(n'' \log\sigma)$ bits. Finally, the $\NBOXED$ array has $m$ entries where $m = (\sigma-1) n'' +1$ and each entry could be upto $n''$. Thus, using the data structure of Theorem~\ref{vector_representation}, $\NBOXED[1..m]$ can be encoded using $(\sigma-1) n'' \log n''+O(\log^2 m)$ bits. Thus, our algorithm produces a representation of the product automaton $\mathcal{P}$ using $(\sigma-1) n'' \log n''+O(n'' \log\sigma)$ bits overall, and this is succinct. This completes the description of the product automaton construction algorithm as stated in Theorem~\ref{thm:product}.

\end{document}